\documentclass[conference]{IEEEtran}
\usepackage{cite}
\usepackage{amsmath,amssymb,amsfonts}
\usepackage{algorithmic}
\usepackage{graphicx}
\usepackage{textcomp}
\usepackage{hyperref}
\usepackage{xcolor}
\usepackage{multirow}

\def\BibTeX{{\rm B\kern-.05em{\sc i\kern-.025em b}\kern-.08em
    T\kern-.1667em\lower.7ex\hbox{E}\kern-.125emX}}

\usepackage{etoolbox}
\makeatletter
\patchcmd{\@makecaption}
  {\scshape}
  {}
  {}
  {}
\makeatother
\usepackage{textcomp}
\newcommand{\textapproxA}{\raisebox{0.5ex}{\texttildelow}}

\begin{document}
\title{Delivering Document Conversion as a Cloud Service with High Throughput and Responsiveness}

\author{\IEEEauthorblockN{1\textsuperscript{st} Christoph Auer}
\IEEEauthorblockA{\textit{IBM Research}\\
Rüschlikon, Switzerland \\
cau@zurich.ibm.com}\\[0.6cm]
\IEEEauthorblockN{4\textsuperscript{th} Cesar Berrospi Ramis}
\IEEEauthorblockA{\textit{IBM Research}\\
Rüschlikon, Switzerland \\
ceb@zurich.ibm.com}
\and
\IEEEauthorblockN{2\textsuperscript{nd} Michele Dolfi}
\IEEEauthorblockA{\textit{IBM Research}\\
Rüschlikon, Switzerland \\
dol@zurich.ibm.com}\\[0.6cm]
\IEEEauthorblockN{5\textsuperscript{th} Peter W.J. Staar}
\IEEEauthorblockA{\textit{IBM Research}\\
Rüschlikon, Switzerland \\
taa@zurich.ibm.com}
\and
\IEEEauthorblockN{3\textsuperscript{rd} André Carvalho}
\IEEEauthorblockA{\textit{SoftINSA Lda.}\\
Tomar, Portugal \\
afecarvalho@gmail.com}\\
}

\maketitle

\begin{abstract}
Document understanding is a key business process in the data-driven economy since documents are central to knowledge discovery and business insights. Converting documents into a machine-processable format is a particular challenge here due to their huge variability in formats and complex structure. 
Accordingly, many algorithms and machine-learning methods emerged to solve particular tasks such as Optical Character Recognition (OCR), layout analysis, table-structure recovery, figure understanding, etc. We observe the adoption of such methods in document understanding solutions offered by all major cloud providers. Yet, publications outlining how such services are designed and optimized to scale in the cloud are scarce. 
In this paper, we focus on the case of document conversion to illustrate the particular challenges of scaling a complex data processing pipeline with a strong reliance on machine-learning methods on cloud infrastructure. Our key objective is to achieve high scalability and responsiveness for different workload profiles in a well-defined resource budget.
We outline the requirements, design, and implementation choices of our document conversion service and reflect on the challenges we faced. Evidence for the scaling behavior and resource efficiency is provided for two alternative workload distribution strategies and deployment configurations. Our best-performing method achieves sustained throughput of over one million PDF pages per hour on 3072 CPU cores across 192 nodes.
\end{abstract}

\begin{IEEEkeywords}
cloud applications,
document understanding,
distributed computing,
artificial intelligence
\end{IEEEkeywords}

\section{Introduction}

\begin{figure*}[!ht]
\center{\includegraphics[width=\textwidth]{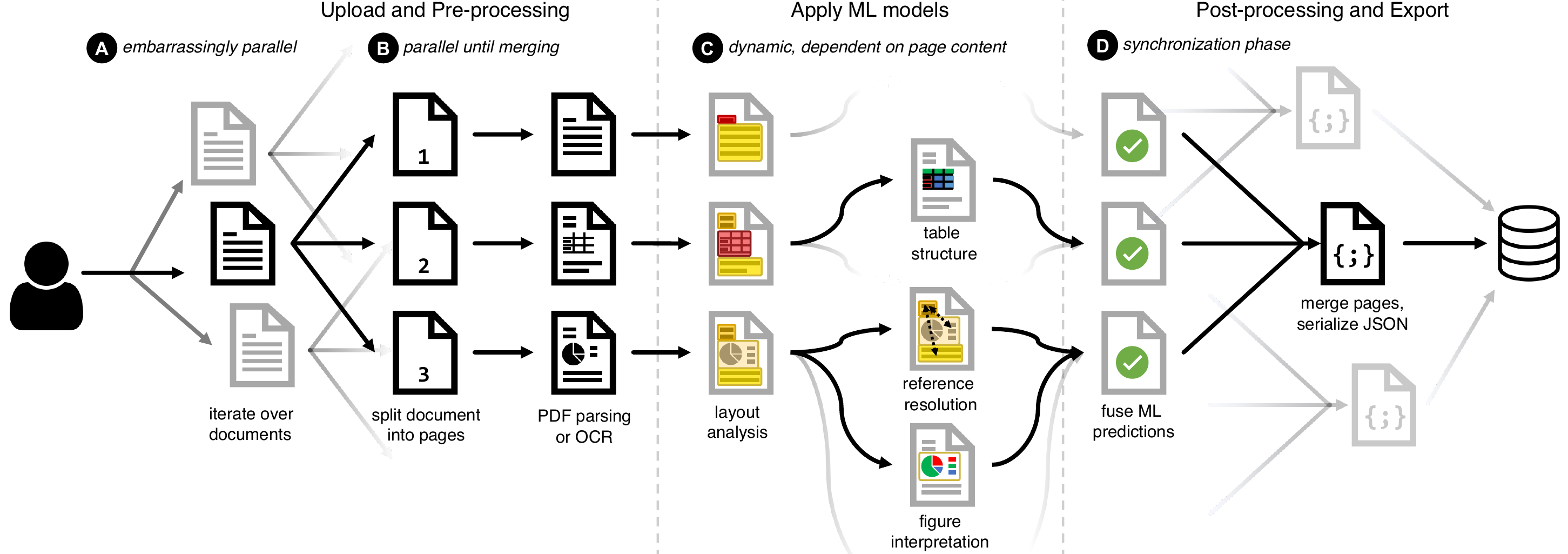}}
\caption{\label{fig:depgraph}
Sketch of operation dependency graph in a document conversion pipeline}
\end{figure*}

Over the past decade, many organizations have accelerated their transformation into data-driven businesses, as studies have shown its positive impact in efficiency, decision making, or financial performance~\cite{newccs:IdcFutureScape2022, newccs:FreeformDynamics2020}. Leading companies are increasingly deploying workloads on public and private cloud infrastructure, including business intelligence processing and machine learning models in data analytics platforms ~\cite{newccs:DataPlatformsMarketMap}. This is owed to several factors such as high availability, lower cost for compute, and storage~\cite{newccs:ForbesCloudAdoption}, as well as the flexibility to scale up or down a cloud-based business process to fit the operational needs. Workloads and services can be containerized, deployed, and orchestrated through widely adopted and standardized platforms like Kubernetes~\cite{newccs:EdcCloudSurvey, newccs:KubernetesBook}.

A key business process relevant to many companies is document understanding. Documents may constitute contracts, guidelines, manuals, presentations, papers, etc., which contain valuable knowledge for their operations. We observe that several specialized companies and all major cloud providers offer dedicated services (SaaS) for various aspects of document understanding such as Optical Character Recognition (OCR) (e.g., Amazon Textract\footnote{https://aws.amazon.com/textract}), forms, and invoice parsing (Docparser\footnote{https://docparser.com}, Nanonets\footnote{https://nanonets.com/invoice-ocr}, Google Document AI\footnote{https://cloud.google.com/document-ai}, Microsoft SharePoint Syntex\footnote{https://docs.microsoft.com/en-us/microsoft-365/contentunderstanding}), or conversion of unstructured formats such as PDF into structured content (IBM Watson Discovery\footnote{https://www.ibm.com/cloud/watson-discovery}).

Conversion of PDF documents into a structured, machine-processable format is a particularly challenging business process due to the high variability and weak normalization of its input. To name a few dimensions of variability, PDF documents can be short or long, encode programmatic or scanned content, have simple or complex page layouts, may contain tables or figures, etc. Thus, the process of recovering their structure and extracting content in high detail entails several dynamic steps (see Fig.~\ref{fig:depgraph}).
On the computational side, this relies on multiple algorithms and machine-learning (ML) models specialized for particular tasks. Examples for such models include OCR~\cite{newccs:lin2020review}, document layout analysis~\cite{newccs:xu2020layoutlm, newccs:doclaynet22, newccs:pagemodelrnn}, table structure recovery~\cite{newccs:zheng2021global, newccs:tableformer2}, figure understanding~\cite{newccs:siegel2018extracting}, reference and citation resolution~\cite{newccs:duma2016rhetorical}, etc.
Furthermore, the ML landscape is evolving rapidly, with new models frequently exposing significantly different characteristics in terms of computational expenses, memory usage, or accelerator requirements.
Additionally, many bottlenecks hide in logical operations such as parsing the PDF code, rasterizing a PDF page to a bitmap, or serialization and database transactions.
On top of that, a cloud service to solve this business task needs to be flexible enough to support different consumption modes. Some users may need to convert large document repositories in bulk, expecting high throughput, while others may want to convert a single document ad-hoc, expecting short response time. Satisfying both expectations in a common application architecture is non-trivial.

In short, a cloud service for document conversion needs to be easily adaptable, must scale with model resource demand and workload sizes, and ideally avoids strong assumptions about the consumption mode. As such, it shares many traits with cloud services for other business processes but also holds unique challenges, which make it an interesting systems engineering target. 
In this paper, we take the subject of document conversion as a case to illustrate the particular challenges of scaling a complex data processing pipeline with strong reliance on ML methods on cloud infrastructure. We outline the practical problems our team faced while developing a cloud service of such nature~\cite{newccs:ccs2018} and present solutions in several aspects. Our main contributions are:

\begin{enumerate}
\item We propose a validated novel approach to efficiently scale an end-to-end processing pipeline leveraging ML in a cloud environment for the use-case of PDF document conversion.
\item We reflect on the design considerations, challenges, and implementation choices of our PDF conversion service.
\item We introduce a set of comparative metrics to evaluate the performance characteristics. 
\item We present and analyze the resulting scaling and response behavior in a benchmark setup on common cloud infrastructure for two alternative workload distribution strategies and deployment configurations.
\end{enumerate}

\section{\label{sec:requirements}Requirements and Objectives for Document Conversion in the Cloud}

As teased in the introduction, a cloud service for document conversion provides a particularly broad and challenging set of requirements, which are not trivial to satisfy all at once. Precisely this combination of real-world requirements make it an interesting target for a case-study on scaling a business process in the cloud. Broadly speaking, we can split the requirements into three categories, namely 1. the ability to process documents with a wide variety in characteristics at scale, 
2. the ability to incorporate and consume several ML models for different tasks and 
3. the ability to perform well in different user consumption modes (\textit{bulk} or \textit{ad-hoc}) which sit on opposite ends. 
It should be noted that only the first category is linked to the domain problem of document conversion. The latter two categories are shared by many cloud services.

\subsubsection*{PDF document processing}
We work under the premise that a document conversion service must treat a single document as an atomic data-unit. For instance, a single output file in JavaScript Object Notation (JSON) needs to be produced for each PDF file that is submitted. Contrary to other data formats such as images or plain text, documents have an internal granularity in the shape of pages. 
This trait can be exploited, since many operations can be executed independently on each page. Examples of such operations are text-cell identification (either through parsing programmatic PDF code or through OCR), layout segmentation, table-structure recovery, figure classification, etc. 
Two critical constraints need to be recognized here. 
Firstly, the intra-page operations may have dependencies. For instance, table-structure recovery relies on text-cell identification and on layout segmentation to locate a table. 
Secondly, the results of each operation on each page needs to be ultimately merged into a single output document (see Figure~\ref{fig:depgraph}, phase D). This merge operation ensures that the content of the document has the correct section-structure and ordering. 
It is therefore an invalid assumption that the conversion of a document can be simply treated as an embarrassingly parallel set of operations over its pages. Rather, to obtain the converted document, one has to execute a complex operation dependency graph which dynamically forks and merges on two levels (see Figure~\ref{fig:depgraph} phase B and C). While this provides opportunity for highly concurrent execution, it can create significant imbalance in the number and runtime of operations for a given page, due to their extremely varying complexity. 
Adding to this, the PDF format has inherent properties that make it very \textit{opaque}. Therefore, there is no practical way to estimate the complexity and resource demand required for processing ahead of execution time. 
From an architecture perspective, the above is suggesting an approach which decomposes the full chain of operations and data into \textit{tasks}. Such tasks can then be created and executed concurrently through an orchestration mechanism that is aware of their dependencies. We outline implementation options and examine the advantages and trade-offs with different task granularity in Sections~\ref{sysdesign} and ~\ref{results}.

\subsubsection*{Machine-learning models}
ML models play a powerful role in the quality of document conversion. 
To date, they out-compete any rule-based algorithm in document conversion tasks and even approach human-like accuracy in particular tasks such as layout analysis~\cite{newccs:binmakhashen2019document}.
From a systems perspective, ML models bring an own set of challenges.
First, solving the conversion task requires many different types of ML models as outlined earlier. Each of these models has different run-time characteristics and resource requirements in terms of CPU cores, memory demand, and accelerators. Furthermore, ML models evolve at a rapid pace.
Consequently, to quickly adopt state-of-the-art ML models in production, our cloud service needs to be flexible with regard to integration of new models. Encapsulation of ML models into separate microservices, each serving on an own endpoint, can provide for both resource isolation and easy integration. We will discuss choices for ML serving and scaling in Section~\ref{sysdesign}.

\subsubsection*{User requirements}
The requirements outlined above are all system-internal requirements. From a user experience perspective, there are strong requirements to be met regarding the conversion speed and the response time. Our goal is to provide a cloud service which supports both bulk-conversion of huge document repositories, as well as ad-hoc conversion of individual documents.
In the former case, a key concern is to achieve the highest (sustained) document throughput and ensure it scales proportionally to the compute resources. 
Importantly, this should allow to set a time-budget for a large conversion workload, e.g., a full online library, and infer the compute resource cost, or conversely, estimate the conversion time with a given resource budget.
In the latter case, the primary objective is to achieve the shortest time-to-solution (TTS), allowing for an interactive user experience. 
Here it is important to note that the smaller a workload is (e.g., one document), the more challenging it becomes to achieve proper workload balancing in a distributed processing scheme.
Both goals are reflecting in the scaling behavior of the system, which we will examine in Section~\ref{scaling}.
Additionally, we want to enable multi-tenant operation for many users on one application instance. Therefore, the service must provide reasonable fairness when working at full capacity and multiple users each submit a conversion job. That means, submitted conversion jobs cannot be simply processed serially, but the service should allow to schedule multiple jobs concurrently with short delay.

\subsubsection*{Evaluation dimensions}
In addition to functional, technical, and experience requirements, we also need to consider performance evaluation metrics, which typically fall into three dimensions: \textit{speed}, \textit{resource cost}, and \textit{quality}. In this paper, we will not focus on conversion \textit{quality}, since it is directly determined by the choice of models and their accuracy. Model performance for different tasks is backed by many peer-reviewed results~\cite{newccs:binmakhashen2019document}. Instead, this paper will focus on \textit{speed} and \textit{resource cost}, and how they correlate given a fixed system configuration with selected state-of-the-art models.

\section{System Design}\label{sysdesign}
Having established the problem domain, requirements, and objectives, this section will outline and reflect on the fundamental application architecture of our document conversion service. Further, we establish a distributed task orchestration mechanism and discuss aspects of its implementation.

\subsection{Application architecture}\label{general_arch}
Two fundamental decisions to be made prior to any implementation are the technology stack and architecture patterns to use. 
Given its clear dominance in virtually every contribution by the ML community, its proven utility in data-science tasks and the broad support in webservice frameworks ~\cite{newccs:statisticstimes2021python, newccs:pythonvsr2021, newccs:jetbrains2021python}, we chose python as the primary implementation language.
Another priority we set is to work close to the bare kubernetes stack, in order to exercise lower-level control over the workload distribution, scaling, and resource creation. For this reason, we ruled out full-fledged data engines (e.g., Apache Spark~\cite{newccs:zaharia2010spark}) and avoided proprietary or vendor-specific tools.

The other critical decision regards the application architecture. Here, one can find fundamentally different options, depending on the type and volume of data, their processing complexity, and the consumption modes. 
As outlined in section~\ref{sec:requirements}, we expect high-volume workloads which can be partitioned horizontally and processed in parallel to a high degree.
On the other hand, we want to satisfy expectations of short response times for ad-hoc conversion with small data volumes.

For the latter case, an architecture optimized for \textit{real-time} processing would appear to be a sensible choice.
An example for this is a microservice-based application with a user-facing, synchronous REST API and a backend which implements the document processing through a chain of remote-procedure-calls (RPC) between these microservices.
With microservice replication, backed by sufficient compute resources or elastic cloud infrastructure, responsive operation could be upheld even for several concurrent users.
Still, this approach would certainly fail to scale to high-volume workloads such as document repositories or libraries, especially in the case where the service is used for ad-hoc conversion and bulk conversion at the same time. Given a bulk conversion workload in the order of hundreds of documents, the client requests would soon stack up and cause congestion, forcing the system to deny more requests. Dynamically, let alone spontaneously, scaling the microservice replication to a level which can absorb this would be extremely challenging in terms of resource management and also be cost-prohibitive.

\begin{figure}[htbp]
\center{\includegraphics[width=\linewidth]{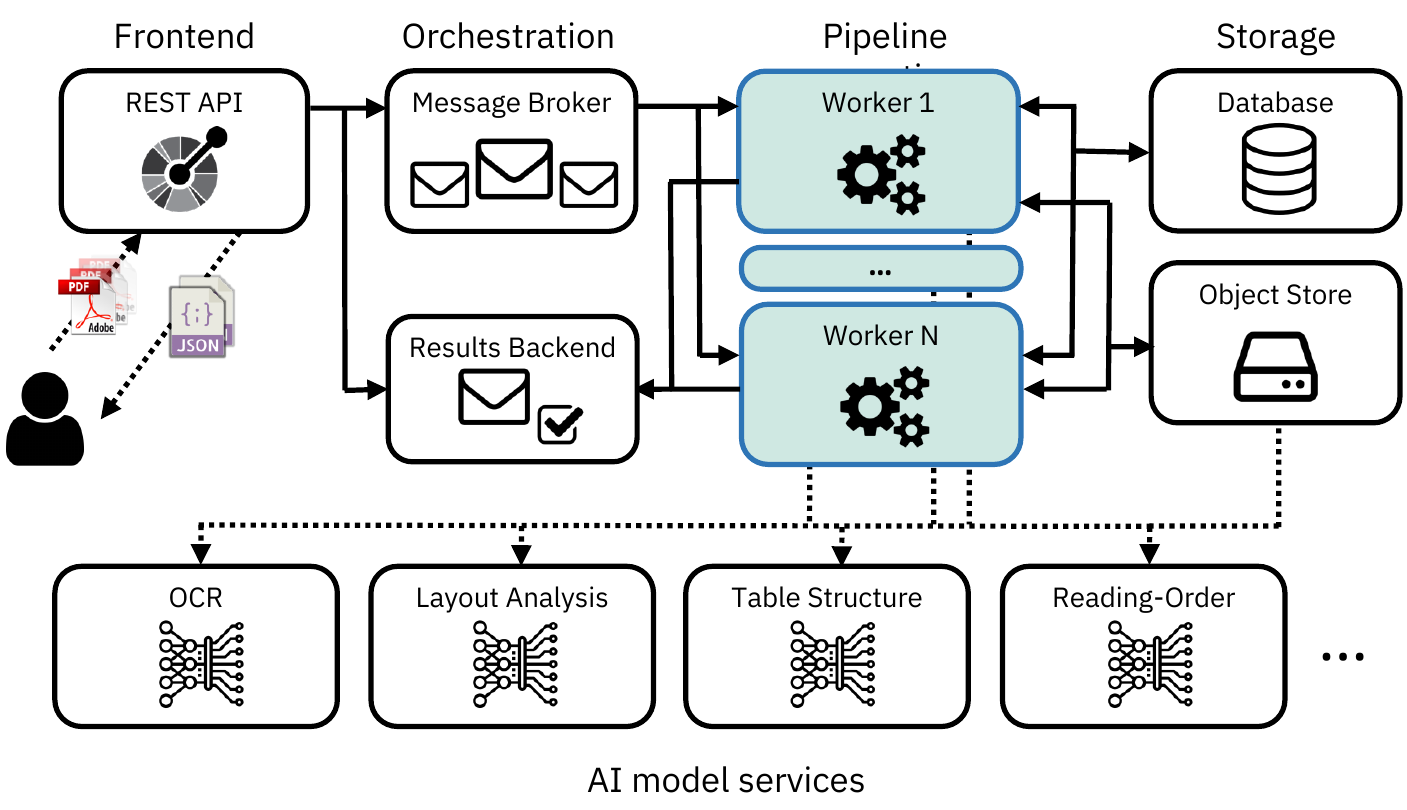}}
\caption{\label{fig:architecture} Architecture diagram of our conversion service. User requests for document conversion are handled by a REST API service, which can dispatch workloads asynchronously to the compute infrastructure through a queueing mechanism. Workers pick up queued tasks from a message broker and store the results for later retrieval on a document database and cloud object storage. The ML models are served through separate microservices, which are consumed by workers when executing the document conversion pipeline.
    }
\end{figure}

For these reasons, we favored an asynchronous task-based \textit{queue-worker} architecture suitable for high-volume batch-processing, involving a task queue and containerized worker processes in the backend. Here, the user-facing API only accepts a document conversion request and enqueues a corresponding task on a message broker. 
Clients receive an immediate response with the task's identifier, which can be used to retrieve the conversion result once the processing is completed. 
The worker processes attach to a task queue, consume tasks competitively, and write status information and task output to a (transient) results-backend and a persistent database (see Figure~\ref{fig:architecture}). 
Since large conversion workloads can be partitioned horizontally, workers are not required to communicate to each other or share resources. Hence, worker processes can be run isolated in containers and trivially replicated both inside and across cluster nodes through kubernetes. 
Tasks produced in a conversion job can reside on the task queue or on a worker process and eventually persist their state on the results-backend. Together, this ensures that a task remains traceable over its whole lifetime.
As such, a queue-worker architecture protects the system from compute congestion through request overload, promises reliability, and enables simple scaling, at the cost of potentially delayed processing.
However, queue mechanisms may also add I/O overhead and latency to the system, even in idle state. Consequently, TTS in ad-hoc conversion may be handicapped.
To minimize such adverse effects, we chose to use \textit{online} workers, which boot at service deployment time. This makes their lifetime independent of any conversion job and allows to pre-load the codebase, initialize resources, and keep internal state across tasks (such as cached connections and database drivers). 
Further, we chose to implement a light-weight library which enables low-latency task orchestration as well as fairness between multiple conversion jobs. Details are provided in the following subsections.

We introduce a notable exception from the queue-worker pattern for serving ML models. Instead of executing ML inference codes inside the workers, we externalize ML models as microservices with a simple REST API. The slightly less efficient interface and added I/O compared to direct code-binding and shared memory inside a worker process has proven to be a reasonable sacrifice to the significantly lower integration effort for new models. Ease of integration is a key requirement in the fast-evolving ML model landscape.
Additionally, model serving can be resource-budgeted and scaled out independently from workers.
The latter is particularly relevant for resource efficiency since our ML inference is computationally far more expensive and at the same time far less dynamic in memory use than other operations (e.g., PDF splitting, parsing, and merging).
Leveraging this benefit would otherwise demand several different worker types and queues (one for each model), therefore increasing the complexity in the core system.
In our service implementation, we wrap ML model codes in a python \textit{aiohttp} webserver runtime and rely on \textit{knative serving} for scaling and load-balancing~\cite{newccs:openshiftservless}. This compares well to the technology used in popular ML serving frameworks such as ray.io~\footnote{https://www.ray.io}, Seldon Core~\footnote{https://www.seldon.io/solutions/open-source-projects/core}, or TensorFlow Serving~\footnote{https://www.tensorflow.org/tfx/guide/serving}.

\subsection{Task orchestration architecture}\label{pipeline}

We define one conversion job as one (asynchronous) call to our service's API endpoint, which may request conversion of any number and length of PDF files. To organize the operations within a conversion job, we define \textit{tasks}. A task is defined by a code function which performs an operation, by the data it processes, and an identifier which makes it traceable across the backend architecture. All tasks belong to a conversion job, which itself acts as the root-task.
Tasks will be executed on the workers and cover all operations, with the exception of ML model inference. 

Obviously, implementing the pipeline outlined in Figure~\ref{fig:depgraph} in a purely serial fashion, meaning as a single task for each conversion job, would not promise any reasonable efficiency, even if multiple jobs were processed in parallel on different workers. 
To use available cluster resources efficiently, we focus on two strategies in particular. First of all, we try to balance the workload distribution evenly. Therefore, we split the conversion jobs into smaller tasks, which can be processed independently on different workers. 
Secondly, we strive to avoid that workers are underutilizing CPU resources while waiting for I/O-bound operations, such as reading and writing files, database transactions, or network requests to the ML models. Therefore, we enable concurrency between multiple tasks on the same worker.

\subsubsection*{Data partitioning and workload distribution}\label{part_workload_distr}

\begin{figure}[htbp]
\center{\includegraphics[width=\linewidth]{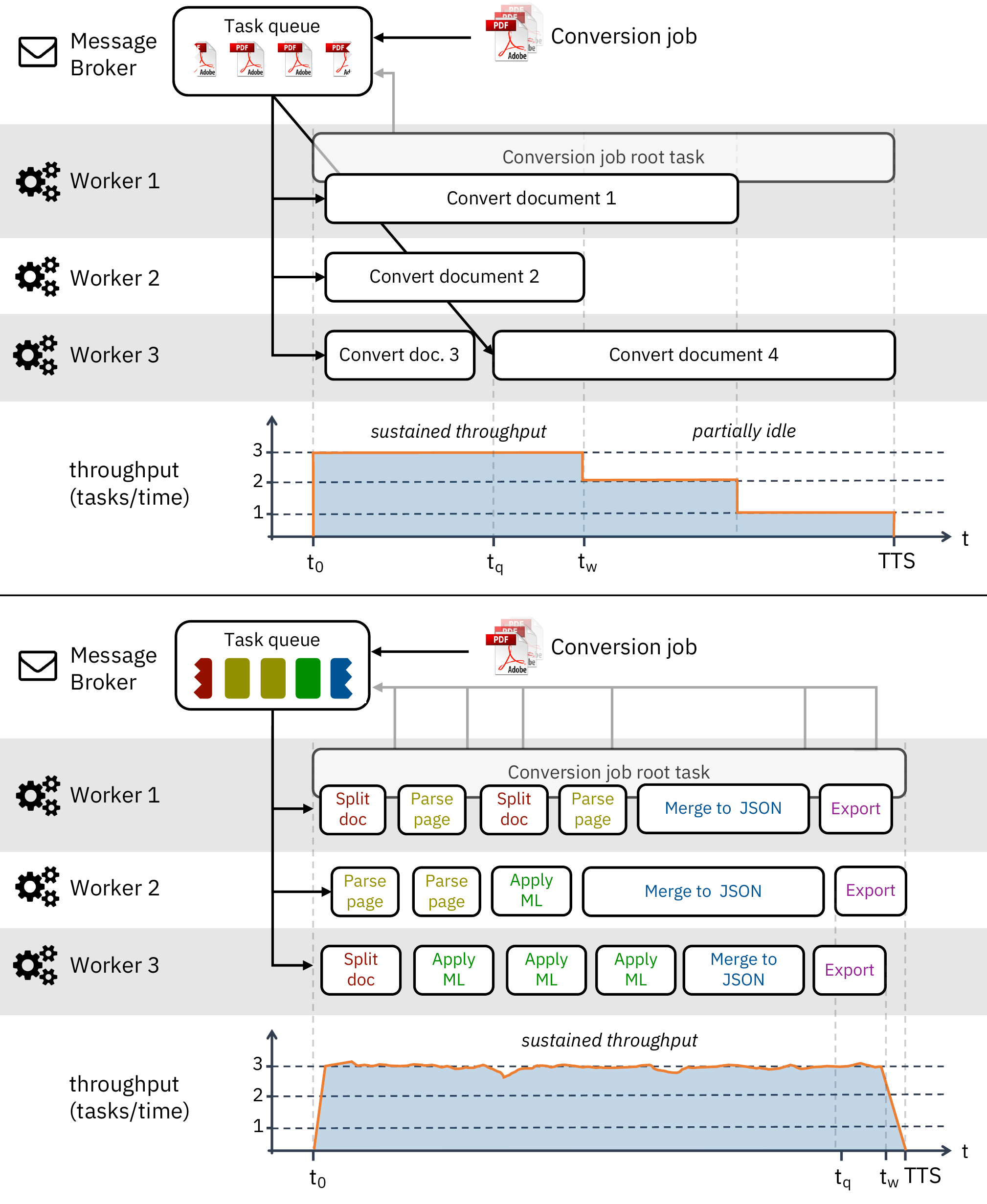}}
\caption{\label{pipeline_tasks}Sketch of workload distribution schemes on the document-level (top) and page-level (bottom). A conversion job is submitted at time t\textsubscript{0}. Tasks are produced and consumed by the workers through the task queue. The queue drains at time t\textsubscript{d}. Sustained throughput is upheld until the first worker finds no more work (t\textsubscript{w}). The job only finishes after completion of all tasks (TTS). }
\end{figure}

We explore two schemes for partitioning data and forming tasks in a conversion job: a) per-document and b) in equal-sized batches of pages, across document boundaries. 
In the former case, each document contained in a conversion job is fully processed into the corresponding JSON output on its own subtask, which we further refer to as the \textit{document-level distribution scheme}.
In the latter case, we further increase granularity by producing tasks for equal-sized batches of pages, further referred to as \textit{page-level distribution scheme} (see Fig.~\ref{pipeline_tasks}). As sketched out, we expect to see stronger workload imbalance in the document-level distribution scheme, leading to underutilization of workers in the tail part of a conversion job (from t\textsubscript{w} until TTS).

A conversion job using the document-level distribution scheme consists of \(N+1\) tasks for \(N\) PDF documents, where the root task has the responsibility of iterating over the data source and submitting one task for every document to the queue. 
While the document-level scheme can be implemented rather straightforward without any dependencies between tasks, the page-level scheme introduces significantly more complexity. It becomes evident from Fig.~\ref{fig:depgraph} that some stages in the pipeline require to work on the full document (e.g., merging pages into the full document JSON output and exporting to the target), while others can be run page-by page (e.g., PDF parsing and applying ML models). We therefore chose to define specialized tasks responsible for each processing stage, which depend on the results of tasks from the preceding stage. 
Hence, many tasks need to be produced on the fly and awaited at multiple points from other tasks. Their count depends on factors determined by the PDF content.
Another consequence is that partial results from each task need to be stored intermediately for downstream tasks to be picked up. Since downstream tasks may schedule on any worker in any node, this incurs additional data transmission to and from a database or object storage. This particular requirement is not present in the document-level pipeline, because all intermediate data can remain worker-local until a document is fully converted. 

\subsubsection*{Task concurrency}\label{task_concurrency}
Keeping allocated cluster resources well saturated, while also not overbooking them, is key to efficient resource usage and high throughput in document conversion.
The operations in the conversion pipeline can be roughly divided into those primarily compute-bound and primarily I/O-bound. Examples for the former are parsing PDF contents, rendering an image of a PDF page, applying ML models or OCR. Examples for the latter are retrieving a PDF file over the network, unpacking an archive, exporting to a target database, or storing and fetching intermediate task results.
A feasible approach we found is to interleave I/O-bound and compute-bound operations on the same worker, while allowing only one compute-bound operation to execute at the same time. Thus, precisely saturating resources on a cluster node turns into a matter of monitoring resource usage of workers under load and adjusting the amount of worker processes deployed on each node accordingly. Concrete evidence of this is presented in section~\ref{experiments}. 

We evaluated and discarded multiple implementation options to achieve interleaving, such as threading and explicit callback-passing code. 
Eventually, we settled on implementing all task code in python \textit{asyncio}\footnote{https://docs.python.org/3/library/asyncio.html} co-routines, which run in an asynchronous event loop.
This cooperative concurrency scheme is more efficient compared to launching and managing threads, but it requires a different coding paradigm. Any active task arriving at code points which require to wait for I/O can choose to suspend and return control to the asyncio event loop, and resume at a later point after its waiting condition has been fulfilled. Compute-intense tasks may fill the waiting time and remain in control of execution until they are completed. As such, no second compute-intense task can compete for resources in the same worker process. 

\subsubsection*{Dynamic task orchestration}\label{distr_orchestration}
Realising the dynamic traits of the proposed task distribution schemes prompted us to find an efficient mechanism to produce (sub-)tasks, submit these to the task queue, and wait for their completion during the pipeline execution.
In the context outlined above, this demands for a task distribution code which can itself run inside an asyncio event-loop, such that in-worker concurrency between tasks can be leveraged.

Several task distribution libraries have emerged in the community for different requirements and use-cases, for instance, ray.io, Celery\footnote{https://docs.celeryq.dev}, Dramatiq\footnote{https://dramatiq.io}, or Apache Beam\footnote{https://beam.apache.org}. To our surprise, none of the available options could meet our needs, since distribution and execution of asynchronous co-routines as tasks is widely unsupported. Few attempts to provide this capability have remained experimental and are poorly maintained or abandoned~\cite{newccs:aiotasks, newccs:celeryPoolAsyncio}.
Therefore, we decided to implement a custom library to orchestrate asynchronous tasks, which we make available as open source under the name \textit{Mognet}\footnote{https://github.com/DS4SD/project-mognet}.
Fundamentally, it allows to create worker processes that initialize an asyncio event loop and attach to an amqp\footnote{https://www.amqp.org} message broker and a redis\footnote{https://redis.io}
instance for storage of every task's status. A worker process subscribes to a queue and fetches up to \(X\) tasks at once, which are then executed on the worker's event loop concurrently. To support the dynamic nature of our pipeline, we explicitly allow any task to produce (many) subtasks and await their completion. Subtasks are submitted to the task queue, like any task. Since all tasks are asynchronous co-routines, waiting for subtasks is a non-blocking and cheap operation.
It is worth mentioning that waiting for subtasks creates a risk for deadlocks. A task may wait for subtasks forever while they cannot be scheduled on any worker, because all workers are occupied up to their limit \(X\). To circumvent this situation, the local limit \(X\) of concurrently scheduling tasks on a worker is incremented by 1 each time a task produces any number of subtasks, and decremented by 1 again when the producing task completes.

Additionally to the orchestration capability outlined above, we exploit the property that such tasks which produce subtasks are very cheap to suspend and reactivate for another significant benefit. Instead of producing and enqueueing every possible subtask at once, the producing task may instead \textit{iteratively} enqueue new subtasks only as previous subtasks finish. This allows precise control over how many subtasks are submitted from a producer task at once and therefore enables to keep the task queue short. 
A concrete example for this case is the root task of a conversion job, which produces one subtask for each document. Given a number of \(N\) workers, enqueueing much more than \(N\) tasks at once is to no benefit for distributed processing. Conversely, it can delay the scheduling of competing conversion jobs. Enqueueing the \(N+1^{st}\) document subtask only when the first document subtask finishes instead gives opportunity for better interleaving of tasks belonging to competing conversion jobs on the queue. The same principle translates to the level of pages inside one document.
We demonstrate the effect of this strategy in section~\ref{fairness}.

\section{Evaluation}\label{results}
To evaluate the real-world performance and behavior of our service, we outline our benchmark environment and characterize a representative test dataset used for conversion. In the following, measurements and observations regarding the scalability and multi-user fairness are presented and discussed.

\subsection{Metrics}\label{metrics_scaling}
Following from the design goals, we establish several evaluation metrics to characterize the scaling behavior.
The time-to-solution (TTS) for a particular conversion job is defined as the time difference between the submission of the first batch of documents and the retrieval of the last converted document.
The throughput is defined as the number of processed pages per unit of time. Notice that the number of processed documents per unit of time is not a good metric as documents have a variable number of pages. The \textit{effective throughput} is computed by determining the number of processed pages from the input dataset and dividing it by the TTS. The \textit{sustained throughput} is defined as the number of processed pages from the input dataset divided by the time in which all compute resources are busy processing (from t\textsubscript{0} until t\textsubscript{w}, see Fig.~\ref{pipeline_tasks}). Importantly, the \textit{effective throughput} will be dependent on the data volume of the conversion job, while the \textit{sustained throughput} is independent of the data volume of the conversion job. In the limit of a very large conversion job, the \textit{effective throughput} will asymptotically reach the \textit{sustained throughput}, as the time proportion of partially idle worker state at the end of a job will be negligible compared to the time in saturated worker state. Additionally, we observe the \textit{serial time} of the benchmark conversion job. It is defined as the sum of all task runtimes in each conversion job and, in the ideal case, it is constant unless a resource bottleneck is present. 

\subsection{Infrastructure}\label{benchmark_setup}

For our benchmark environment, we created a managed OpenShift cluster (version 4.8, using kubernetes 1.21) hosted on IBM Cloud with default settings and we added a pool of virtualized dedicated cluster nodes. Each cluster node provides 16 CPU cores and 16 GB RAM.
This choice of nodes falls well within the band of optimal resources-to-cost ratio on most cloud infrastructure providers.
To enable ML model serving, we installed \textit{knative serving} through the OpenShift serverless operator~\cite{newccs:openshiftservless}.

Data blobs (i.e., source documents and conversion output) and intermediate results (e.g., model predictions) are stored on managed instances of IBM Cloud Object Storage (COS) and IBM Cloud MongoDB, respectively. 
All resources are hosted in an IBM Cloud location in Frankfurt, Germany. 

\subsection{Test Dataset}\label{datasets}

\begin{figure}[htb]
\center{\includegraphics[width=.9\linewidth]{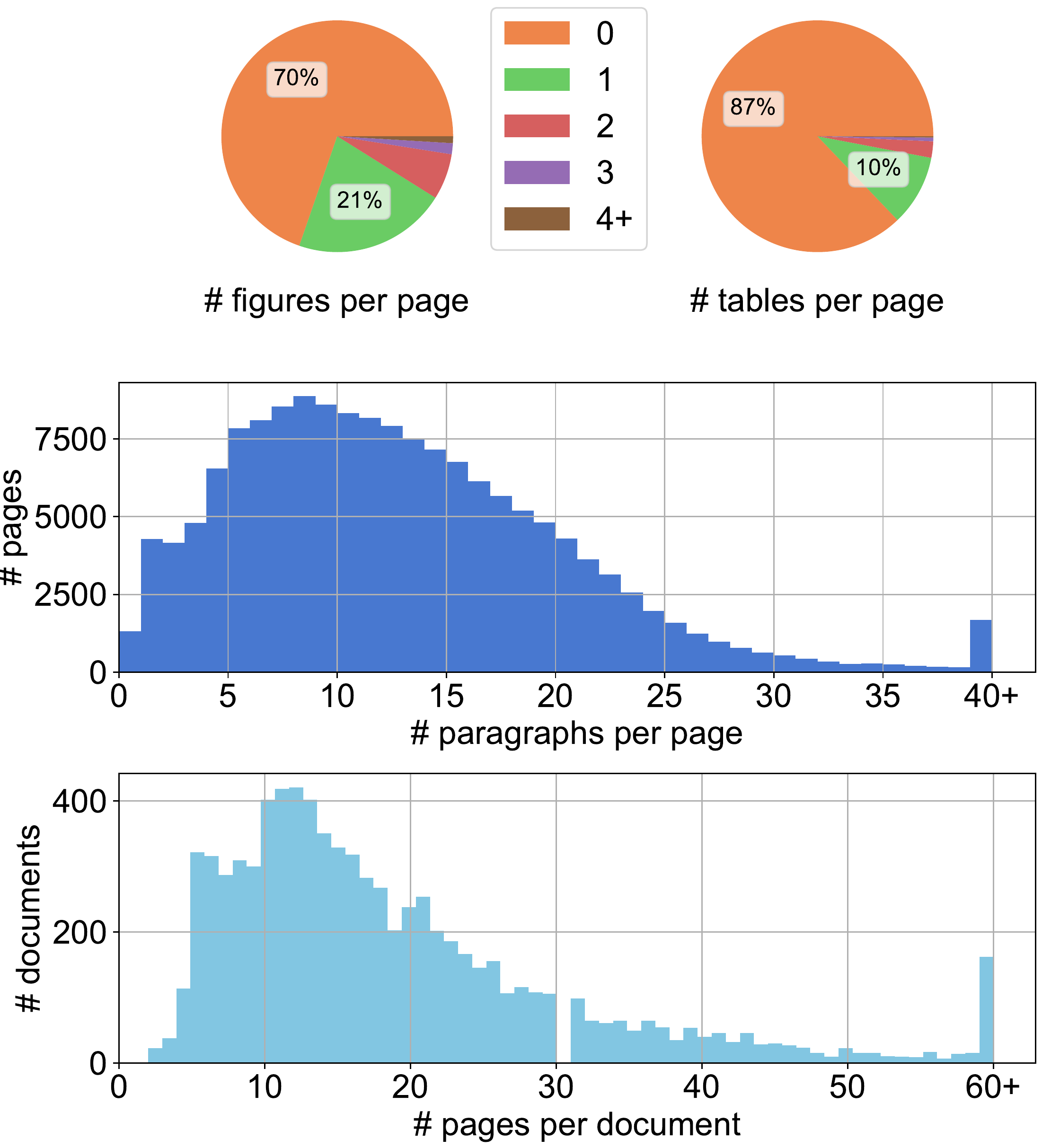}}
\caption{\label{dataset} Characteristics of the test dataset.
}
\end{figure}

To obtain a realistic test dataset, we compiled a subset of 8053 PDF documents from arXiv (consecutive submissions 2106.00001 to 2106.08072, omitting those that do not provide a PDF). It contains 156~529 pages and measures 20.7 GB in binary size. We picked this dataset as it is available at minimal cost\footnote{https://arxiv.org/help/bulk\_data\_s3} and fully reproducible. Furthermore, the arXiv is a representative dataset, as it is one of the main open-source libraries for disseminating scientific preprint articles. Detailed characteristics of the test dataset are shown in Fig.~\ref{dataset}. 

It is worth noting that we did not cherry-pick the documents in the dataset in any way to best reflect a real-world situation. We  have observed that a fraction of documents reproducibly fail to convert or cause outliers in task runtimes. Three documents consistently fail to convert because of PDF interpretation issues. In another 160 documents, a total of 504 pages could not be converted due to memory constraints or rendering problems, which is a typical fraction that we also observed with other types of datasets.

\subsection{Application setup}\label{experiments}

We evaluate two different profiles for the deployment and configuration of our system components to understand the impact of concurrent task interleaving on resource efficiency. For both profiles, we take the resource budget of a cluster node as the basis. 
To decide how many pod replicas of workers and ML models one can safely schedule on a cluster node without overbooking resources, we elected their average memory footprint under load as the primary criterion. 
Our target was to occupy 75\% of the node memory, leaving 25\% as overhead. This choice was made since memory, in contrast to CPU cores, cannot be oversubscribed and therefore strongly determines the required cost.
Average memory footprint and CPU usage of the components, observed under load, are shown in table~\ref{tab:deployresources}.

\bgroup
\def\arraystretch{1.1}
\begin{table}[htb!]
\centering
\caption{\label{tab:deployresources}Typical resource usage of workers and ML models under load in deployment profiles A and B.}
\begin{tabular}{ll|lll}
                           &                  & \multicolumn{2}{c}{component} &  \\
                           &                  & worker         & ML model     &  \\ \cline{1-4}
\multirow{3}{*}{Profile A} & Memory           & 450 MB         & 500 MB       &  \\
                           & CPU              & 0.4 cores     & 0.7 cores   &  \\
                           & Replica ratio    & 1              & 1            &  \\ 
                           \cline{1-4}
\multirow{3}{*}{Profile B} & Memory           & 700 MB         & 500 MB       &  \\
                           & CPU              & 1.2 cores    & 0.7 cores   &  \\
                           & Replica ratio    & 1              & 4            &  \\ 
\end{tabular}
\end{table}
\egroup

In profile A, we simulate a service which is not capable of interleaving concurrent tasks in the same worker process. Thus, we create equal counts of worker replicas and ML model replicas, and effectively disable worker-internal task concurrency by fetching only one task at a time from the queue. We schedule 17-18 pods (8-9 each for worker and ML model) per cluster node within the given memory budget.

In profile B, we intend to optimize the conversion throughput. With the prior knowledge that ML model inference is the computationally most expensive operation in the pipeline, we create worker replicas and ML model replicas at a ratio of 1:4, and allow four concurrent tasks per worker to interleave. This allows to fit 15 pods (3 workers, 12 ML models) per cluster node, since now workers require more memory on average to interleave four tasks at a time (see table ~\ref{tab:deployresources}).

Further, we do not enforce limits on CPU usage per worker or ML model replica through kubernetes. Worker processes are naturally bounded to a single core due to python's global interpreter lock ~\cite{newccs:beazley2010understanding}. Each replica of the ML models is configured to use only one thread to avoid CPU congestion.

As for the ML models, we deliberately set up a rather expensive object-detection network (Faster R-CNN with ResNet50 backbone, \textapproxA{63M} weights)~\cite{newccs:detectron2modelzoo} for layout segmentation, which we feed PDF pages rasterized to 1025$\times$1025 pixels.
Bare inference time using one CPU core is in the order of 3 seconds. The structure of identified tables is reconstructed by a secondary model, here using a very cheap topological algorithm with negligible resource demand.

\subsection{Scalability}\label{scaling}

To evaluate the scaling behavior of the service, we measure the runtime and throughput of exactly one conversion job at a time, repeated with increasing node pool sizes and matching application deployment scale on the benchmark environment outlined above.

\subsubsection*{Benchmark series and protocol}\label{bench_protocol_scaling}
All evaluation metrics defined above are measured repeatedly in both deployment profiles after scaling the pool size to 12, 48, 96, and 192 cluster nodes (using 192, 768, 1536, and 3072 CPU cores). The scaling level of 12 cluster nodes acts as the baseline measurement to determine the speedup for the higher scales.
On a second dimension, the whole test series is carried out separately for both task distribution schemes (document-level and page-level).

\begin{figure*}[htb!]
\center{\includegraphics[width=\textwidth]{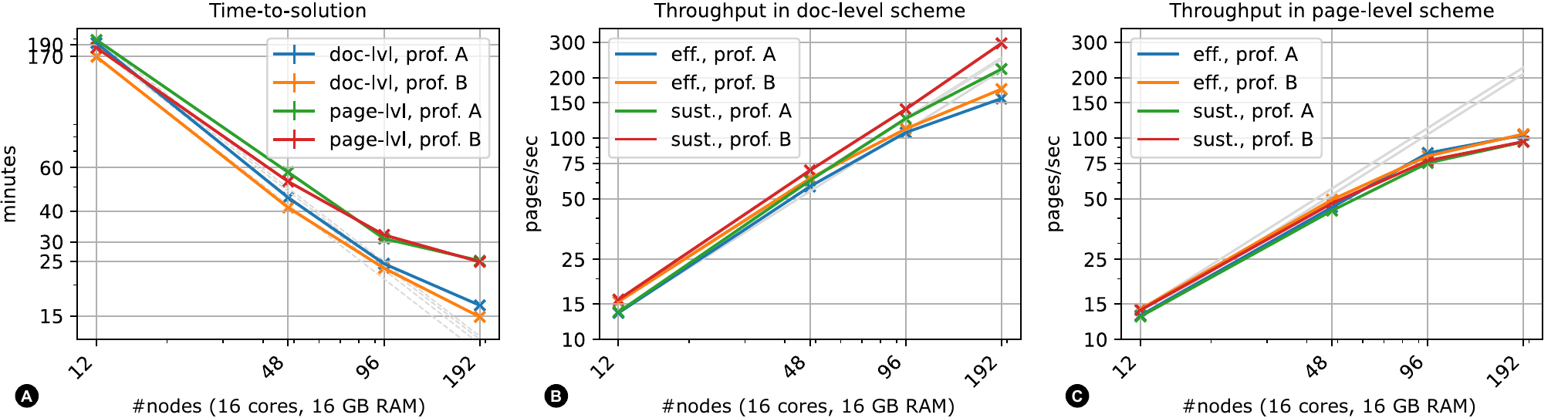}}
\caption{\label{fig:scaling} Scaling of time-to-solution (A), effective throughput and sustained throughput on document-level distribution scheme (B), page-level distribution scheme (C) in both deployment profiles with node count. Gray diagonals indicate theoretical linear scaling.}
\end{figure*}

\subsubsection*{Observations in scaling behavior}\label{scaling_results}

\begin{figure}[b!]
\center{\includegraphics[width=0.8\linewidth]{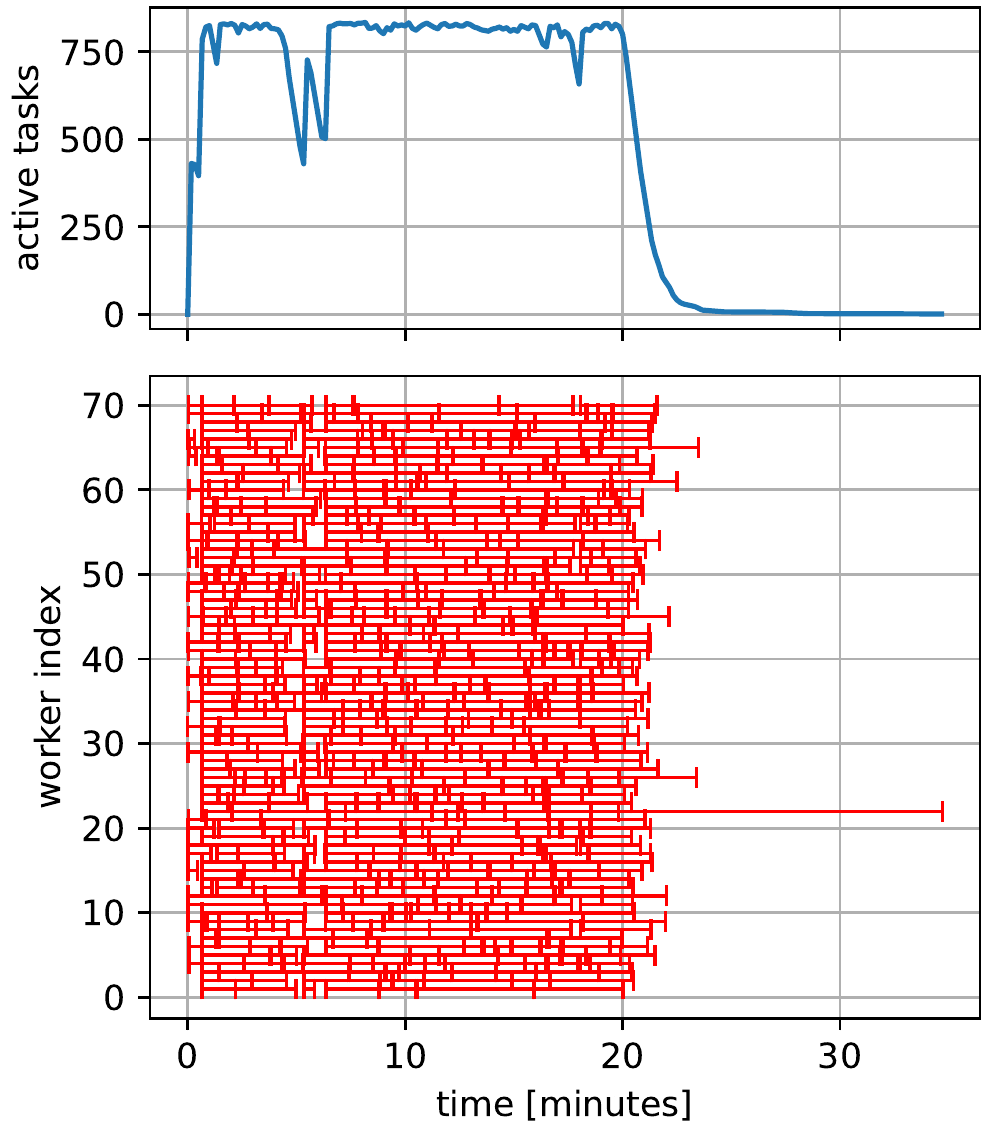}}
\caption{\label{fig:task_inspection_profile} Active tasks over job runtime (top) and time schedule in 70 of 834 workers (bottom), using document-level scheme, profile A on 96 nodes. Each red bar marks the runtime of one task, equal to one document. The job completes only 14 minutes after the first worker finds no more work due to few, long documents in the tail. An anomaly around 7 min is caused by worker restarts. }
\end{figure}

Several observations can be made in the context of our scaling benchmark series, as shown in Fig.~\ref{fig:scaling}. First of all, we find that the TTS at the lowest scale (12 nodes) is very similar across both deployment profiles and task distribution schemes.
As we scale up the number of cluster nodes, we first recognize that both distribution schemes achieve reasonable speedup in TTS.
However, the TTS in the document-level distribution scheme beats the TTS in page-level distribution scheme by a growing margin at increasing scales. This observation also mirrors in the effective throughput. 
At the largest scale (192 nodes), the achieved speedup in TTS starts to tail off to different degrees in both schemes, irrespective of the deployment profile. Our test dataset converts within 17 minutes in the best case (document-level scheme, profile B) and 25 min in the worst case (page-level scheme, profile A). Contrary to our theoretical considerations, the loss in scaling efficiency affects the page-level scheme much stronger than the document-level scheme. 
Deeper investigation revealed that two different causes are responsible for the observed loss in scaling efficiency. 
In the document-level distribution scheme, long documents increasingly dominate the TTS of the full conversion job, some of which take over 14 minutes to convert alone (see Fig~\ref{fig:task_inspection_profile}). Thus, TTS would be lower-bounded at 14 minutes even with infinite resources.

In the page-level distribution scheme, despite the fine-grained workload balancing achieved through equal-sized batches of pages, we find that the amount of I/O created by storing and loading intermediate task results creates a scaling bottleneck. Our MongoDB instance cannot keep up with the concurrent transaction load. This bottleneck is completely absent in the document-level distribution scheme since it's implementation creates no database transactions at all. 
Further, interleaving tasks on the same worker (profile B) does not demonstrate significant benefit to TTS or effective throughput over the non-interleaving setup (profile A) in either of the distribution schemes. 
A different picture shows when looking at the sustained throughput.
In the document-level distribution scheme, we are observing even slightly better-than-linear scaling, which is clearly distinct from the effective throughput. At first glance, this artifact may seem questionable. However, deeper verification shows two contributing factors to this. The first is owed to small imperfections in the load-balancing of requests to our ML model replicas. The fewer ML model replicas exist, the more pronounced is the impact of an unbalanced distribution of requests to those. Relatively speaking, a larger fraction of requests will be subject to additional wait-times for the response.
The second factor is a peculiarity of our worker implementation. To regularly clean up the excess memory allocated (but not used) in the python runtime, worker pods reboot after every 64 completed (document-level) tasks. This causes a delay of approximately 10 seconds each time. In each conversion job, such reboots occur more frequently when fewer workers need to consume more tasks. Therefore, we see sustained throughput improve overproportionally at higher scales.

The significant difference between sustained and effective throughput in the document-level distribution scheme is clearly explained through the workload imbalance between workers, as shown in figure~\ref{fig:task_inspection_profile}.
Moreover, we find that deployment profile B, which exploits in-worker task concurrency, achieves consistently higher sustained throughput than profile A with the same total amount of resources. At the highest scale of 192 nodes, sustained throughput of 296 pages/second is achieved in profile B, compared to 220 pages/second in profile A.
In the page-level distribution scheme, effective and sustained throughput remain much closer together and follow similar curves across scale levels. While this indicates better workload balancing, both metrics are equally impacted by the I/O bottleneck described before.

\begin{figure}[htb]
\center{\includegraphics[width=0.9\linewidth]{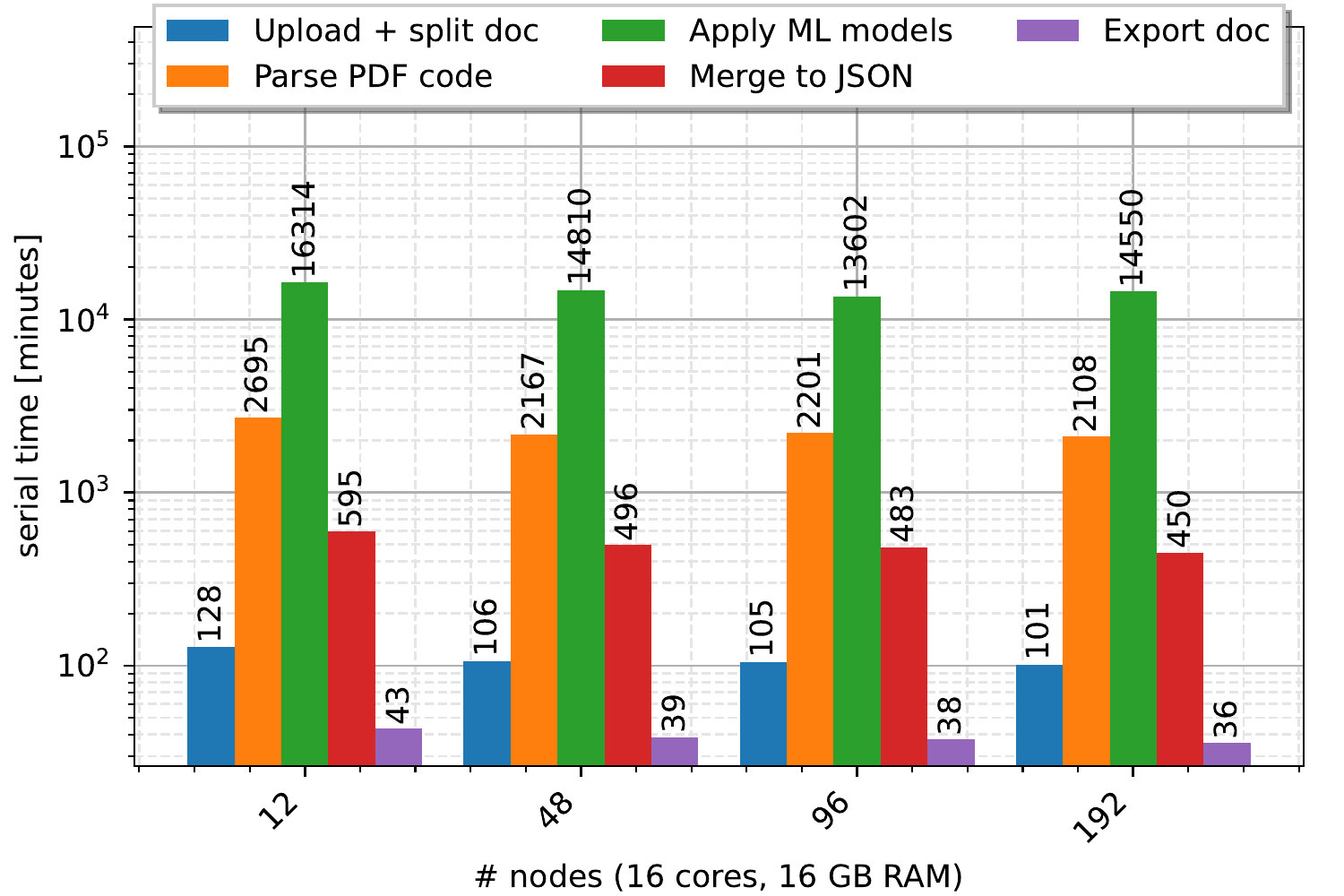}}
\caption{\label{serial_time} Measured serial times of operations in the document-level pipeline,  profile A, across all scales.}
\end{figure}

An important context for the interpretation of the observations above is provided by the serial time, differentiated by processing stage (see figure~\ref{serial_time}). 
We see consistent serial times on every tested scaling level in the document pipeline, which confirms that no resource congestion or bandwidth limitation is present. In the page-level pipeline (not shown), serial times, in particular for upload and merge operations, increase noticeably on the largest scale.
Finally, we clearly recognize that applying ML models is the most time consuming operation, responsible for approximately 76\% (page-level scheme) to 84\% (document-level scheme) of the task runtimes in our test conversion job. 

\subsection{Fairness}\label{fairness}
To understand both how fast and how fair our service handles multiple concurrent conversion jobs, we analyze an extreme case. While saturating our service with the conversion of our benchmark dataset, we submit a second conversion job with a single average-length document (15 pages) one, two, and three minutes later. We measure the TTS of the single-document job repeatedly and compare it to the TTS of the same job submitted on an idle system. This experiment is carried out at a scale of 48 nodes and repeated for both deployment profiles and task distribution schemes. Results are shown in table~\ref{tab:fairness}.

\bgroup
\def\arraystretch{1.3}
\begin{table}[]
\centering
\caption{\label{tab:fairness}TTS for a 15-page document in idle and busy system states. Values represent mean \textpm\ standard deviation in seconds.}
\begin{tabular}{clll}
\multicolumn{1}{l}{}                    &                                   & \multicolumn{2}{c}{distribution scheme}                    \\
\multicolumn{1}{l|}{profile}            & \multicolumn{1}{l|}{system state} & \multicolumn{1}{l|}{document-level}    & page-level        \\ \hline
\multicolumn{1}{c|}{\multirow{2}{*}{A}} & \multicolumn{1}{l|}{idle}         & \multicolumn{1}{l|}{32.6 \textpm\ 1.1}   & 26.3 \textpm\ 0.9   \\
\multicolumn{1}{c|}{}                   & \multicolumn{1}{l|}{busy}         & \multicolumn{1}{l|}{114.9 \textpm\ 17.5} & 248.1 \textpm\ 17.3 \\ \hline
\multicolumn{1}{c|}{\multirow{2}{*}{B}} & \multicolumn{1}{l|}{idle}         & \multicolumn{1}{l|}{32.6 \textpm\ 0.2}   & 25.7 \textpm\ 0.7   \\
\multicolumn{1}{c|}{}                   & \multicolumn{1}{l|}{busy}         & \multicolumn{1}{l|}{109.6 \textpm\ 47.7} & 220.4 \textpm\ 34.2
\end{tabular}
\end{table}
\egroup

The single document converts with a reproducible TTS of 33s (document-level scheme) or 26s (page-level scheme). On a busy system, this increases by a factor of 3.5 (document-level scheme) or 7.5 (page-level scheme) on average, with a substantial standard deviation between 30\% and 45\% of the TTS.

One can draw multiple conclusions from these results.
First, we see good evidence that our strategy to keep the length of the task queue short (see section~\ref{distr_orchestration}) pays off. Concurrent conversion workloads of very different volume are processed in a reasonably fair manner, all while avoiding the complexity of job priorities or separate queues. Still, the TTS on a busy system is significantly higher in the page-level distribution scheme, since the single-document conversion job produces a total of 12 tasks, which pass through the task queue, compared to only one in the document-level distribution scheme. 

Second, we find that the page-level distribution scheme on this 15-page document yields 20\% lower TTS (1.25x speedup) than the document-level scheme in an idle system. Disregarding all non-parallelizable operations (upload, merge, and export), the theoretical speedup would be bounded at 4x for 15 pages, because we chose to configure our system to create tasks for batches of up to four pages. Dynamically deciding the optimum page-batch size based on the page count could improve this aspect.
In reality, we must consider two factors: On one hand, profiling the page-level distribution scheme reveals that approximately one quarter of the total conversion job run-time is spent in the non-parallelizable operations for this document. On the other hand, the page-level distribution pays a high price on additional I/O, as shown earlier. In the observed case, it requires 12 distinct store and retrieve operations, while the document-level scheme requires only two. As such, TTS in ad-hoc conversion can profit from the page-level distribution scheme only with longer documents.

Third, we find that the deployment profile (A or B) has no relevant impact on the measurements above. Processing 15 pages only utilizes a fraction of the available worker and ML model replicas here, which explains why no difference is observed.

\section{Conclusions}

We have carefully established a scalable cloud service design for the task of document conversion and presented the impact of different workload distribution schemes and implementation options on both absolute speed and scaling behavior. 
With the best-performing setup, we achieve a sustained conversion throughput scaling linearly with the resource budget up to 192 nodes (equal to 3072 CPU cores and 3072 GB of memory). In practice, this allows us to convert over one million PDF pages per hour. 

An important lesson we learned is that distributing the workload more evenly at a page-level cannot compensate for the excess of I/O expenses and synchronization need it creates over document-level distribution. In terms of sustainable throughput, it ultimately steers into a scaling bottleneck. More efficient intermediate data storage and locality may push this bottleneck further up the scale, but it will eventually prevail. We find good potential in using page-level distribution, however, to achieve short response time in ad-hoc conversion of few documents.

To drive a production-grade deployment of our conversion service, we need to consider some important differences to the benchmark setup presented here. For cost reasons, it is highly advisable to rely on automatic scaling mechanisms.
The ML model microservices can be easily transferred to a managed platform-as-a-service where the replicas can be dynamically scaled according to the request load without pre-allocating cluster infrastructure (e.g., Google App Engine\footnote{https://cloud.google.com/appengine}, IBM Cloud Code Engine\footnote{https://www.ibm.com/cloud/code-engine}, or Amazon ECS\footnote{https://aws.amazon.com/ecs}). 
Additionally, worker pods and even cluster nodes may be automatically added and removed based on the length of the task queue. 
In deployments for open audiences, we enforce strict timeouts on tasks to ensure that very heavy or even poisonous PDF files are not locking up resources forever.

The investment of implementing all pipeline logic in asynchronous co-routines and developing a task distribution library which works natively within an event-loop has proven to be worthwhile. 
Distributing and executing code with dynamic subtasks and dependencies becomes very cheap and it helps to achieve fairness in multi-tenant usage with little effort and complexity. Additionally, it allows to write idiomatic code.
It should be noted however that the efficiency advantage of interleaving tasks as proposed here depends significantly both on the amount of opportunity to avoid blocking I/O and on the proportion of compute-bound code executing in the workers. 
Externalizing the ML models as web-services has proven to be a sensible choice in this regard.
Considering the large fraction of total compute-time spent in ML inference, investing effort into cheaper ML models will remain a top priority in our on-going work. 

In the landscape of data-driven business, several applications share important properties with document conversion. One such case is automated knowledge-base construction, where several data processing pipelines have been developed for extraction, normalization, and analysis of knowledge~\cite{newccs:staar2020corpus}, often targeting specific domains like material sciences, geology, or biomedicine~\cite{newccs:jacobsson2022open, newccs:pyzer2022accelerating, newccs:piantanida2021using, newccs:badal2019challenges,DBLP:journals/corr/abs-1907-08400}. They commonly process complex or unstructured input data in bulk (e.g., literature, experiment reports, and instrument data), employ machine-learning (e.g., language understanding), and often require complex dynamic pipelines. As such, the insights presented and conclusions drawn in this paper would be of value for the design and validation of future work on scalable processing pipelines in the cloud. 

\bibliographystyle{bibtex/bst/IEEEtran}
\bibliography{bibtex/bib/IEEEabrv,bibtex/bib/newccs}

\end{document}